\def\mbar{\overline{m}}
\def\half{{\textstyle{{1}\over{2}}}}
\def\frak#1#2{{\textstyle{{#1}\over{#2}}}}
\def\frakk#1#2{{{#1}\over{#2}}}

\def\GeV{{\rm GeV}}
\def\eV{{\rm eV}}
\def\Ocal{{\cal O}}
\def\Lcal{{\cal L}}
\def\nutil{\tilde \nu}
\def\Ytil{\tilde Y}
\def\Btil{\tilde B}
\def\Qtil{\tilde Q}
\def\Ltil{\tilde L}
\def\etil{\tilde e}
\def\util{\tilde u}
\def\dtil{\tilde d}
\def\sy{supersymmetry}
\def\DRED{\hbox{DRED}}
\def\NSVZ{\hbox{NSVZ}}
\def\DREDp{\hbox{DRED}'}
\def\npb{{Nucl.\ Phys.\ }{\bf B}}
\def\plb{{Phys.\ Lett.\ }{ \bf B}}

\def\prd{{Phys.\ Rev.\ }{\bf D}}

\def\prl{Phys.\ Rev.\ Lett.\ }

\def\sic{supersymmetric}

\def\pa{\partial}

\def \inparg{\leftskip = 40pt\rightskip = 40pt}
\def \outparg{\leftskip = 0 pt\rightskip = 0pt}
\def\msbar{{\overline{\rm MS}}}
\def\drbar{{\overline{\rm DR}}}
\input harvmac
\input epsf
\input tables
{\nopagenumbers
\line{\hfil LTH 654}
\line{\hfil hep-th/0505238}
\line{\hfil Revised Version}
\vskip .5in
\centerline{\titlefont Two-loop $\beta$-functions and their effects} 
\medskip
\centerline{\titlefont for the R-parity Violating MSSM}
\vskip 1in
\centerline{\bf I.~Jack, D.R.T.~Jones and A.F.~Kord}
\bigskip
\centerline{\it Department of Mathematical Sciences,  
University of Liverpool, Liverpool L69 3BX, U.K.}
\vskip .3in

We present the full two-loop $\beta$-functions for the MSSM including 
R-parity violating couplings.
We analyse the effect of two-loop running on
the bounds on R-parity violating couplings, on the nature of the LSP
and on the stop masses. 

\Date{May 2005}}
\newsec{Introduction}
The minimal \sic\ standard model (MSSM) consists of a supersymmetric
extension of the standard model, with the addition of a number  of
dimension 2 and dimension 3 \sy-breaking mass and interaction terms.
It is well known that  the MSSM is not, in fact,
the most general renormalisable field theory consistent
with the requirements of gauge invariance and naturalness;
the unbroken theory is augmented by a discrete symmetry
($R$-parity) to forbid a set of baryon-number and lepton-number
violating interactions, and the \sy-breaking sector omits
both $R$-parity violating soft terms and a set
of ``non-standard'' (NS) soft breaking terms. There is a large literature
on the effect of R-parity violation; a recent analysis
(with ``standard'' soft-breaking terms) and references
appears in Ref.~\ref\ADD{B.C.~Allanach, A.~Dedes and H.K.~Dreiner,
\prd69 (2004) 115002}; for earlier relevant work see in particular
\ref\deCarlosDU{
B.~de Carlos and P.L.~White,
\prd 54 (1996)  3427;  {\it ibid\/} 55 (1997)  4222
}. The need to consider NS terms  in a model--independent
analysis was stressed in
Ref.~\ref\lrlh{L.J. Hall and  L. Randall,
\prl 65 (1990) 2939};
for a discussion of the NS terms both in general
and in the MSSM context see Ref.~\ref\bfpt{F.~Borzumati, G.R.~Farrar, 
N.~Polonsky and
S.~Thomas, \npb555 (1999) 53}\nref\jj{I.~Jack and D.R.T.~Jones,  
\plb 457 (1999)  101}\nref\hethr{J.P.J.~Hetherington,
JHEP  0110 (2001)  024}\nref\jjka{I.~Jack, D.R.T.~Jones and A.F.~Kord,
\plb588 (2004) 127}--\ref\demir{D.A.~Demir, G.L.~Kane and T.T.~Wang, 
\prd72 (2005) 015012}; however in this paper we shall ignore the NS terms. 

The unification of the three gauge couplings in the MSSM at a scale of around
$M_X\sim10^{16}\GeV$ provides compelling evidence both for  
supersymmetry and for the existence of an underlying unified theory. 
We shall consider the standard mSUGRA scenario where we assume just three
parameters at the unification scale, namely universal scalar and gaugino masses,
and a universal trilinear scalar coupling, $m_0$, $m_{1/2}$ and $A$ 
respectively. The remaining parameters are $\tan\beta$ and sgn($\mu)$. The 
complete mass spectrum is determined by running the couplings and masses from 
$M_X$ to $M_Z$ (taking the quark masses and gauge couplings at $M_Z$ as 
additional inputs). There is an extensive literature on this process in
the R-parity conserving (RPC) case and some pioneering work in the R-parity 
violating (RPV) case. In particular, Ref.~\ADD\ contained a comprehensive 
analysis of RPV effects on various scenarios, using full one-loop
$\beta$-functions for RPV parameters, and additionally including 2-loop RPC 
corrections for RPC parameters. Our purpose here is to make available the full
2-loop $\beta$-functions for both RPC and RPV parameters and to explore the
effect of incorporating these full $\beta$-functions on a representative sample
of the scenarios considered in Ref.~\ADD; in particular neutrino masses and the
nature of the lightest supersymmetric particle (LSP). 
We shall not actually present the
$\beta$-functions explicitly but rather refer the reader to a website 
\ref\web{http://www.liv.ac.uk/$\sim$dij/rpvbetas/} where
they can be accessed for the most general case (including a general $3\times3$ 
matrix of Yukawa couplings).
   
\newsec{The Soft $\beta$-functions}
For a  general $N=1$ \sic\ gauge theory with superpotential
\eqn\newW{
 W (\phi) = \frak{1}{2}{\mu}^{ij}\phi_i\phi_j + \frak{1}{6}Y^{ijk}
\phi_i\phi_j\phi_k,}
the standard soft \sy-breaking scalar terms are
as follows
\eqn\newV{\eqalign{V_{\hbox{soft}} &=
\left(\frak{1}{2}b^{ij}\phi_i\phi_j
+ \frak{1}{6}h^{ijk}\phi_i\phi_j\phi_k +\hbox{c.c.}\right)
+(m^2)^i{}_j\phi_i\phi^j,\cr}}
where we denote $\phi^i \equiv \phi_i^*$ etc.

The complete exact results for the soft $\beta$-functions
are given by\ref\jjpa{I.~Jack and   D.R.T.~Jones
\plb415 (1997) 383 }%
\nref\jjpb{I.~Jack, D.R.T.~Jones and A.~Pickering,
\plb432 (1997) 114}
--\ref\akk{L.V.~Avdeev, D.I.~Kazakov and I.N.~Kondrashuk,
\npb510 (1998) 289}:
\eqn\allbetas{\eqalign{
\beta_M &= 2\Ocal\left[\frakk{\beta_g}{g}\right],\cr
\beta_{h}^{ijk} &= h{}^{l(jk}\gamma^{i)}{}_l -
2Y^{l(jk}\gamma_1{}^{i)}{}_l, \cr
\beta_{b}^{ij} &=
b{}^{l(i}\gamma^{j)}{}_l-2\mu{}^{l(i}\gamma_1{}^{j)}{}_l,\cr
\left(\beta_{m^2}\right){}^i{}_j &= \Delta\gamma^i{}_j,\cr}}
where $\gamma$ is the matter multiplet anomalous dimension, and
\eqna\Otdef$$\eqalignno{
{\cal O}  &= Mg^2\frakk{\partial}{\partial g^2}-h^{lmn}  
\frakk{\partial}{\partial Y^{lmn}},
&\Otdef a\cr
(\gamma_1)^i{}_j  &= {\cal O}\gamma^i{}_j,
&\Otdef b\cr
\Delta &= 2\Ocal\Ocal^* +2MM^* g^2{\partial
\over{\partial g^2}}
+\left[\Ytil^{lmn}{\partial\over{\partial Y^{lmn}}}
+ \hbox{c.c.}\right]
+X{\partial\over{\partial g}}.&\Otdef c\cr
}$$
Here $M$ is the gaugino mass and
$\Ytil^{ijk} = (m^2)^i{}_lY^{jkl} +  (m^2)^j{}_lY^{ikl} + (m^2)^k{}_lY^{ijl}.$
Eq.~\allbetas{}\ holds in a class of renormalisation schemes that includes
$\DRED'$\ref\jjmvy
{I.~Jack, D.R.T~Jones, S.P.~Martin, M.T.~Vaughn and Y.~Yamada,
Phys.\ Rev.\ D {\bf 50}, 5481 (1994)
}, which we will use throughout. Finally the $X$ function above is given   
(in the NSVZ scheme
\ref\jnsvz{D.R.T.~Jones, \plb123 (1983)  45 \semi
V.~Novikov et al, \npb229 (1983) 381  \semi
V.~Novikov et al, \plb166 (1986) 329 \semi
M.~Shifman and A.~Vainstein, \npb277 (1986)  456}) by
\eqn\exX{
X^{\NSVZ}=-2{g^3\over{16\pi^2}}
{S\over{\left[1-2g^2 C(G)(16\pi^2)^{-1}\right]}}}
where
\eqn\Awc{
S =  r^{-1}\tr [m^2C(R)] -MM^* C(G),}   
$C(R),C(G)$ being the quadratic Casimirs for the matter and adjoint
representations respectively.
There is no corresponding exact form
for $X$ in  the $\DREDp$ scheme\jjmvy;
however we only require here the leading contribution which is the same in
both schemes; the subleading $\DREDp$ contribution is given in 
Ref.~\ref\jjpb{I.~Jack, D.R.T.~Jones and A.~Pickering,
\plb 432 (1998) 114}. These formulae can readily be specialised to the case 
of the RPV MSSM and their implementation can be automated; in our case we 
used the FORM package. (We have also implemented this procedure up to three 
loops for the RPC MSSM\ref\jjkb{I.~Jack, D.R.T~Jones and A.F.~Kord,
\plb579 (2004) 180\semi
Ann. Phys. 316 (2005) 213}, and made the results available on another 
website\ref\webtwo{http://www.liv.ac.uk/$\sim$dij/betas}.)

In our analysis we also include ``tadpole'' contributions,
corresponding to renormalisation of the Fayet-Iliopoulos (FI) $D$-term
at one and two loops.
These contributions are not expressible exactly in terms of
$\beta_{g_i}, \gamma$; for a discussion see
Ref.~\ref\jjp{I.~Jack and D.R.T.~Jones, \plb473 (2000) 102\semi
I.~Jack, D.R.T.~Jones and S.~Parsons, \prd62 (2000) 125022\semi
I.~Jack and D.R.T.~Jones, \prd63 (2001) 075010}. For
universal boundary conditions, the FI term is very small at
low energies if it is zero at gauge unification.

\newsec{The R-parity Violating MSSM}
The unbroken ${\cal N} = 1$ theory is defined 
by the superpotential
\eqn\superpot{W = W_1 + W_2,}
where
\eqn\sperrpc{
W_1 =  Y_u Q  u^c H_2 +   Y_d Q d^c  H_1 +  Y_{e} L  e^c  H_1 +\mu H_1H_2}
 and
\eqn\sperrpv{
W_2 = \frak{1}{2} (\Lambda_E) e^c L L +  \frak{1}{2}(\Lambda_U) u^c d^c d^c
+ (\Lambda_D) d^c L Q +\kappa_iL_iH_2. }
In these equations, generation $(i,j\cdots)$, $SU_2 (a,b\cdots)$,
and $SU_3 (I, J\cdots)$  indices are
contracted in ``natural'' fashion from left to right, thus for example
\eqn\indics{
 \Lambda_D d^c L Q
\equiv  \epsilon_{ab}(\Lambda_D)^{ijk} (d^c)_{iI} L^a_{j} Q^{bI}_k
.}
For the generation indices we indicate complex conjugation by 
lowering the indices, thus $(Y_u)_{ij} = (Y^*_u)^{ij}$. 

We now add soft-breaking terms as follows:
\eqn\spersspc{\eqalign{
L_1 &= \sum_{\phi}
m_{\phi}^2\phi^*\phi + \left[m_3^2 H_1
H_2 + \sum_{i=1}^3\half M_i\lambda_i\lambda_i  + {\rm h.c. }\right]\cr
&+ \left[ h_u Q  u^c H_2 +  h_d Q  d^c  H_1 +  h_{e} L  e^c  H_1
+ {\rm h.c. }\right],\cr
L_2 &= m_R^2   H_1^* L + m_K^2 L H_2 + \frak{1}{2} h_E e^c L L   
+  \frak{1}{2} h_U u^c d^c d^c
+ h_D d^c L Q  + {\rm h.c.}\cr}}
We shall also use the notation
\eqn\ldef{
\lambda^{ijk}\equiv\left(\Lambda_E\right)^{kij},\quad
\lambda^{\prime ijk}\equiv\left(\Lambda_D\right)^{kij},\quad
\lambda^{\prime\prime ijk}\equiv\left(\Lambda_U\right)^{ijk},}
with $h$, $h'$ and $h''$ defined similarly in terms of $h_E$, $h_D$
and $h_U$ respectively. Note that
\eqn\lsymm{
\lambda^{jik}=-\lambda^{ijk},\quad \lambda''^{ikj}=-\lambda''^{ijk},}
with similar symmetry properties for $h$ and $h''$. 

It can be convenient 
to define $\Lcal^a_{\alpha=0\ldots3}=\{H_1^a,L^a_{i=1,2,3}\}$. The 
couplings $\lambda_{\alpha\beta k}$, $\lambda'_{ij\alpha}$ are then 
defined so as to subsume $\Lambda_E$, $Y_e$ and $\Lambda_D$, $Y_d$
respectively; i.e $\lambda_{i0k}=Y_{eik}$, $\lambda'_{ij0}=-Y_{dij}$.
$h_{\alpha\beta k}$, $h'_{ij\alpha}$ are defined similarly. In the same 
spirit we define $\mu_{\alpha}=\{\mu,\kappa_i\}$ and
$b_{\alpha}=\{m_3^2, \left(m_K^2\right)_i\}$; and finally $m_{\Lcal}^2$
incorporates $m_L^2$, $m_R^2$ and $m_{H_1}^2$. 

\newsec{RGE Running and the Mass Spectrum}
The $\drbar$ dimensionless couplings at $M_Z$ are determined from the 
$\msbar$ gauge couplings and the physical quark masses by incorporating 
\sic\ threshold corrections. The boundary conditions on the soft parameters 
and masses are imposed at the unification scale $M_X$. As mentioned earlier
we adopt mSUGRA boundary conditions at $M_X$, so we take     
\eqn\bcs{\eqalign{
m_{\Qtil}(M_X)=m_{\util}(M_X)=&m_{\dtil}(M_X)=m_{\Ltil}(M_X)=m_{\etil}(M_X)
=m_01,\cr
m_{H_1}=&m_{H_2}=m_0,\cr}}
where $1$ is the $3\times3$ unit matrix in flavour space.
\eqn\bcsa{\eqalign{
\kappa_i(M_X)=&(m_R^2)_i(M_X)=(m_K^2)_i(M_X)=0,\cr
M_1(M_X)=&M_2(M_X)=M_3(M_X)=m_{\frak12}.\cr}}
Finally we define 
\eqn\bcsb{\eqalign{
h_u(M_X)=&A_0Y_u(M_X),\qquad h_d(M_X)=A_0Y_d(M_X), \qquad
h_e(M_X)=A_0Y_e(M_X),\cr
h_U(M_X)=&A_0\Lambda_U(M_X),\qquad 
h_D(M_X)=A_0\Lambda_D(M_X), \qquad h_E(M_X)=A_0\Lambda_E(M_X).\cr}}

After running all the couplings from $M_X$ to $M_Z$, the sparticle 
spectrum can be computed. Because of the interdependence of the 
boundary conditions at $M_Z$ and $M_X$ (the threshold corrections
depend on the sparticle spectrum; the unification scale depends on the
dimensionless couplings) we determine the couplings by an iterative
process, reimposing the respective boundary conditions at each iteration.
We define gauge unification to be the scale where $\alpha_1$ and $\alpha_2$ 
meet; we speed up the determination of this by (at each iteration) adjusting
the unification scale using the solution of the one-loop $\beta$-functions
for the gauge couplings from the previous value of the scale.
We employ one-loop radiative corrections as detailed in
Ref.~\ref\pbmz{
D.M.~Pierce, J.A.~Bagger, K.T.~Matchev and R.J.~Zhang,
\npb 491 (1997) 3
}. A particular subtlety in the RPV case is 
that the RGE evolution of $\kappa$ depends on $\mu$, and that of
$m_K^2$ on $\mu$ and $\Btil$. Therefore it is not sufficient (as in the 
RPC case) to determine $\mu(M_Z)$ and $\Btil(M_Z)$ after the iteration, 
from the electroweak breaking conditions; rather, $\mu$ and $\Btil$ must be  
included in the iteration process to establish values of $\mu(M_X)$ and 
$\Btil(M_X)$ which are compatible with the other boundary conditions. A 
second complication in the RPV case is the possibility of sneutrino vevs
$v_i$, 
which satisfy 
\eqn\vdef{
v^2=v_u^2+v_d^2+\sum_{i=1}^3v_i^2={2M_W^2\over{g_2^2}},}
where $v_{d,u}$ are the $H_{1,2}$ vevs, $\tan\beta$ is defined as usual to be
\eqn\tandef{
\tan\beta={v_u\over{v_d}}}
and with our conventions $v=174\GeV$. Then at each iteration, 
$\mu(M_Z)$ and $\Btil(M_Z)$ are determined from\ADD 
\eqn\musqB{\eqalign{
|\mu|^2=&{\left[\mbar^2_{H_1}+(m_R^2)_i{v_i\over{v_d}}
+\kappa_i^*\mu{v_i\over{v_d}}\right]-\left[\mbar^2_{H_2}
+|\kappa_i|^2-{1\over2}(g^2+g_2^2)v_i^2-(m_K^2)_i{v_i\over{v_u}}\right]
\tan^2\beta\over{\tan^2\beta-1}}\cr
&-{1\over2}M_Z^2,\cr
\Btil=&{\sin2\beta\over2}\Bigl\{\left[\mbar^2_{H_1}+\mbar^2_{H_2}
+2|\mu|^2+|\kappa_i|^2\right]\cr
&+\left[(m_R^2)_i+\kappa^*_i\mu\right]{v_i\over{v_d}}-(m_K^2)_i{v_i\over{v_u}}
\Bigr\},}}
where
\eqn\mbdef{\eqalign{
\mbar^2_{H_2}=&m^2_{H_2}+{1\over{2v_u}}{\pa\Delta V\over{\pa v_u}},\cr
\mbar^2_{H_1}=&m^2_{H_1}+{1\over{2v_d}}{\pa\Delta V\over{\pa v_d}},\cr}}
with $\Delta V$ being the one-loop corrections to the scalar potential (we 
assume the sneutrino vevs are real).
Next the sneutrino vevs may be determined from  
\eqn\vmat{
(M_{\nutil}^2)_{ij}v_j=-\left[(m_R^2)_i+\mu^*\kappa_i\right]v_d
+(m_K^2)_iv_u-{1\over2}{\pa\Delta V\over{\pa v_i}},}
where
\eqn\vmata{\eqalign{
(M_{\nutil}^2)_{ij}=&(m_L^2)_{ji}+\kappa_i\kappa^*_j+\frak12M_Z^2\cos2\beta
\delta_{ij}\cr
&+{g^2+g_2^2\over2}\sin^2\beta(v^2-v_u^2-v_d^2)\delta_{ij}.\cr}}
Here $g$ is the $U_1$ electroweak coupling (usually written $g'$).
The one-loop corrections to the effective potential for the RPV MSSM which 
appear in ${\pa\Delta V\over{\pa v_i}}$ were obtained from Ref.~\ref\chunkang{
E.J.~Chun and S.K.~Kang, \prd61 (2000) 075012}. We have included the
squark contributions from Ref.~\chunkang, correcting an obvious typo (a missing 
``$\ln$''); the next most significant corrections, from
charged slepton/Higgs,
given there seem clearly wrong on dimensional grounds and we have omitted 
them; they are much smaller in any case. If (as we do in the neutrino mass
calculation) we impose electroweak
symmetry breaking at the supersymmetry scale $M_{\rm{SUSY}}$ (defined here
as the geometric mean of the stop masses) then the effect even of the 
squark contributions from Ref.~\chunkang\ is negligible. For 
${\pa\Delta V\over{\pa v_{u,d}}}$ we have used the RPC corrections given in
Ref.~\pbmz. (For the calculations of selectron, stau and stop masses given later
we incorporate one-loop threshold corrections and therefore the choice of 
EWSB scale should be less significant; and in fact we choose to evaluate 
the sparticle masses at their own scale.)  
 
Our philosophy throughout is to investigate qualitative effects, particularly
of using two-loop rather than one-loop $\beta$-functions. Therefore we have
made various simplifications in our procedures. ADD consider three standard
forms for the relation between the weak-current and quark-mass bases for the 
couplings, where there is either no mixing, or the mixing is all in the 
down-quark sector, or all in the up-quark sector. We have assumed that the 
Yukawa matrices are diagonal in the weak-current basis both at the GUT scale 
and at the weak scale. This corresponds to assuming a trivial 
CKM matrix, $V_{\hbox{CKM}}=1$ at the weak scale. We are also neglecting the 
generation of off-diagonal Yukawa couplings in the evolution from $M_Z$ to
$M_X$ (an effect which we believe is negligible to the accuracy at which
we are working). 

\newsec{Neutrino Masses}
Here we set bounds on the couplings $\lambda$, $\lambda'$ from the 
cosmological neutrino bound. Combining the 2dFGRS data\ref\colless{M.~Colless
et al, astro-ph/0306581}\ with the WMAP 
measurement\ref\WMAP{D.~Spergel et al, Astrophys. J. Suppl. 148 (2003) 175}\
one gets a bound on the neutrino mass
\eqn\bound{
\sum_im_{\nu_i}<0.71\eV.}
The neutrino mass is given by
\eqn\neutmass{
m_{\nu}={\mu(M_1g_2^2+M_2g^2)\sum_{i=1}^3\Lambda_i^2\over{
2\left(v_uv_d(M_1g_2^2+M_2g^2)-\mu M_1M_2\right)}},}
where
\eqn\Lamdef{
\Lambda_i=v_i-v_d{\kappa_i\over{\mu}}.}
A single non-zero RPV coupling at $M_X$ will generate non-zero $\kappa$,
$m_R^2$ and $m_K^2$ leading to a non-zero neutrino mass. We follow Ref.~\ADD\
in choosing the SPS1a mSUGRA point, which has the following parameter
values at $M_X$:
\eqn\spsa{
m_0=100\GeV,\quad m_{\frak12}=250\GeV,\quad A_0=-100\GeV
\quad \tan\beta=10, \quad \hbox{sign}(\mu)=+. }
Eq.~\bound\ then leads to an
upper bound on the given RPV coupling. 
We assume that only one out of the set of couplings 
\eqn\couplingset{
S_{\lambda}=\{\lambda'_{333}, \lambda'_{322},
\lambda'_{311}, \lambda_{233}, \lambda_{232},\lambda_{131}\}}
is non-zero at $M_Z$, and that only these couplings are non-zero in the 
running; these very nearly form a closed set in any case,
since the only additional couplings which could be generated (at one loop) are 
$\lambda'_{211}$, $\lambda'_{222}$ and $\lambda_{121}$. 
Looking at the form of the $\beta$-functions one can see that these 
couplings could not in any case be generated at a level close to their 
limiting values, since the coupling in $S_{\lambda}$ responsible for 
generating them has a much smaller limiting value and is additionally
suppressed by small (1st or 2nd generation) RPC Yukawa couplings.
Moreover (if we start with just one of them non-zero)
these couplings do not generate any off-diagonal contributions to 
$Y_{u,d,e}$ so our assumption about the form of these matrices at $M_X$
is justified.
 
The bounds on these couplings are shown in Table~1.  
\bigskip
\vbox{
\begintable
Coupling| 1 loop | 2 loop RPC | 2 loop \cr
$\lambda'_{333}(M_Z)$ |$1.0\times10^{-5}$ |$8.7\times10^{-6}$
|$8.4\times10^{-6}$ \cr
$\lambda'_{322}(M_Z)$ |$4.0\times10^{-4}$ |$3.4\times10^{-4}$ | 
$3.2\times10^{-4}$\cr
$\lambda'_{311}(M_Z)$ |$7.0\times10^{-3}$ |$5.9\times10^{-3}$ |$5.6\times10^{-3}$ \cr
$\lambda_{233}(M_Z)$ |$6.5\times10^{-5}$ |$5.3\times10^{-5}$ |$5.4\times10^{-5}$ \cr
$\lambda_{232}(M_Z)$ |$1.1\times10^{-3}$ |$1.0\times10^{-3}$ |$9.2\times10^{-4}$\cr 
$\lambda_{131}(M_Z)$ |$2.2\times10^{-1}$|$1.9\times10^{-1}$|$1.8\times10^{-1}$
\endtable}
\centerline{{\it Table~1:\/} Upper bounds on $\lambda(M_Z)$, $\lambda'(M_Z)$}
\medskip
The ``2 loop RPC'' column
corresponds to the procedure followed in Ref.~\ADD, where the full one-loop
$\beta$-functions were used and also the two-loop RPC corrections were included 
in the $\beta$-functions for the RPC couplings and masses.
Our results (in the 2-loop RPC case) 
agree well with those of ADD, particularly for the $\lambda'$ limits
where we agree to better than $2\%$.

\newsec{The Nature of the LSP}
In the RPV case the LSP is no longer stable and therefore no longer subject 
to cosmological constraints on stable relics. Also the LSP need not be
electrically and colour neutral. Once again we follow Ref.~\ADD\ in taking 
for this analysis the case of ``no-scale'' supergravity, which corresponds 
to taking $A_0=m_0=0$. We shall consider the variation of the nature of the LSP
with $\lambda_{231}$. In this case the LSP is either a stau or a selectron.
The computation of selectron masses is in general more complex than the 
RPC case, since the charged Higgs mix with the charged sleptons, giving 
mass terms of the form 
\eqn\higslep{
{\cal L}_{\hbox{ch}}=-\pmatrix{h_2^- &\etil_{L\gamma}&\etil_{Rk}\cr}
M^2_{\hbox{ch}}\pmatrix{h_2^+\cr \etil^*_{L\delta}\cr\etil^*_{Rl}\cr},}
where $M^2_{\hbox{ch}}$ is an $8\times8$ matrix given by
\eqn\Mchar{
M^2_{\hbox{ch}}
=\pmatrix{(m^2)_{11}+D&b^*_{\delta}+D_{\delta}&\lambda_{\beta\alpha l}\mu^*
v_{\beta}\cr
b_{\gamma}+D^*_{\gamma}&(m^2)_{\delta\gamma}+\lambda_{\alpha\gamma l}
\lambda_{\beta\delta l}v_{\alpha}v_{\beta}+D_{\gamma\delta}&
h_{\alpha\gamma l}v_{\alpha}-\lambda_{\alpha\gamma l}\mu^*_{\alpha}v_u\cr
\lambda^*_{\beta\alpha k}\mu_{\alpha}v_{\beta}&h^*_{\alpha\delta k}v_{\alpha}
-\lambda_{\alpha\delta k}\mu_{\alpha}v_u&
(m^2_E)_{lk}+\lambda_{\alpha\beta k}\lambda_{\alpha\gamma l}v_{\beta}v_{\gamma}
+D_{lk}\cr},}
where
\eqn\mdefs{\eqalign{
(m^2)_{11}=&\mbar^2_{H_2}+|\mu_{\alpha}|^2,\cr
D=&\frak14(g_2^2+g^2)(v_u^2-\sum_{\alpha}v_{\alpha}^2)
+\frak12g_2^2v_{\alpha}^2,\cr
D_{\delta}=&\frak12g_2^2v_uv_{\delta},\cr
(m^2)_{\gamma\delta}=&\left(m_{\Lcal}^2\right)_{\delta\gamma}
+\mu_{\gamma}\mu^*_{\delta},\cr
D_{\gamma\delta}=&\frak14(g_2^2-g^2)(v_u^2-\sum_{\alpha}v_{\alpha}^2)
\delta_{\delta\gamma}+\frak12g_2^2v_{\gamma}v_{\delta},\cr
D_{lk}=&\frak12g^2(v_u^2-\sum_{\alpha}v_{\alpha}^2)\delta_{lk}.\cr}}
However if $\lambda_{231}$ is the only RPV coupling, the matrix is still 
diagonal except for the standard stau mixing. 
We have included the one-loop corrections to the slepton masses as given in 
Ref.~\ref\pbmz{
D.M.~Pierce, J.A.~Bagger, K.T.~Matchev and R.J.~Zhang,
\npb 491 (1997) 3
}. Of course these omit any corrections from RPV couplings but presumably
these will be extremely small. 

In Fig.~1 we show the variation of the nature of the  
LSP with $\tan\beta$ and $\lambda_{231}(M_X)$. Here we have used the two-loop
RG evolution equations but in fact the results using one-loop evolution are
almost identical. Moreover it is easy to check that (at least at one loop)
if $\lambda_{231}$ is the only non-zero RPV coupling at $M_Z$ then it will 
remain so at all scales, so we can use a simplified set of 
$\beta$-functions in which we only retain $\lambda_{231}$.

\bigskip
\epsfysize= 4in
\centerline{\epsfbox{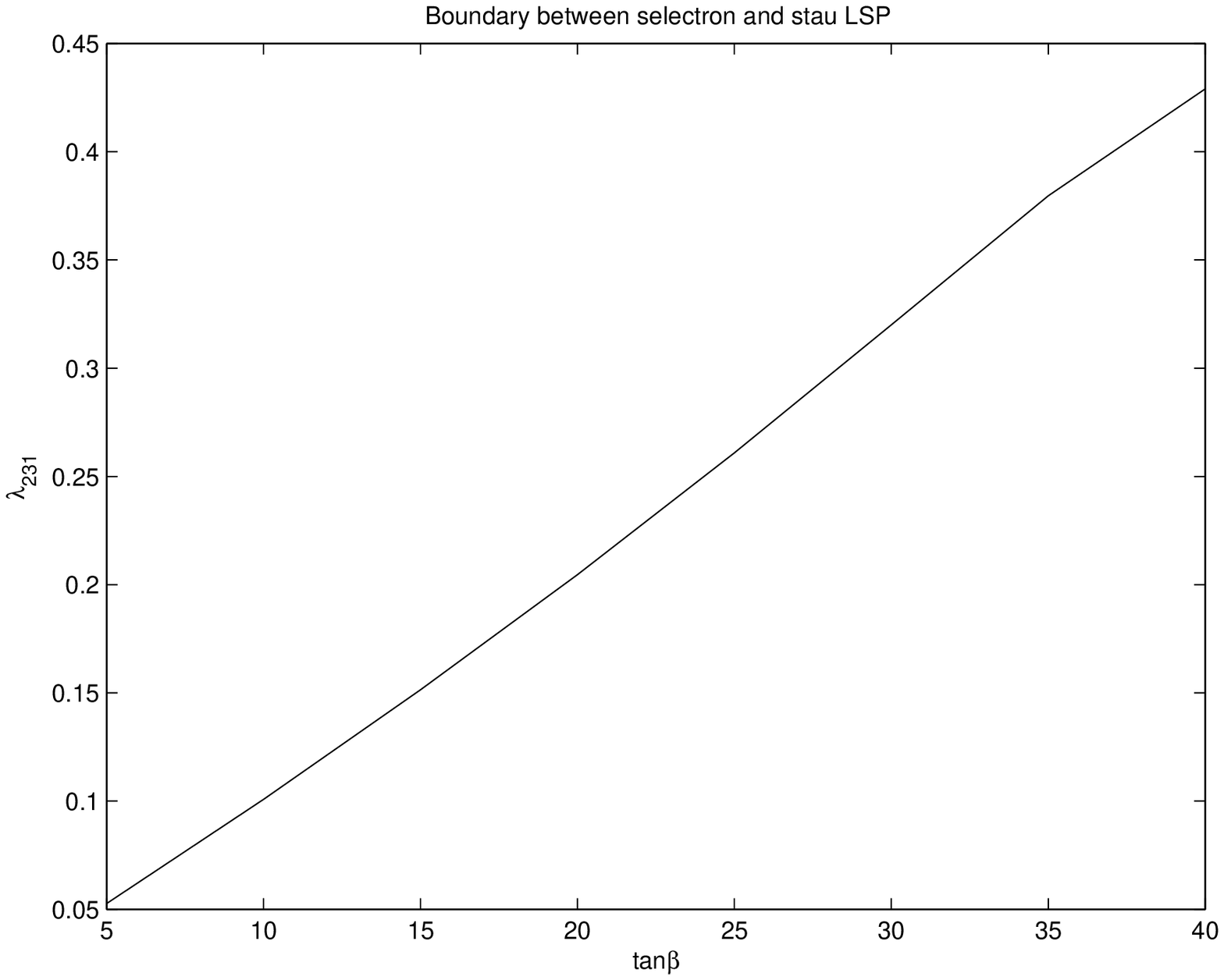}}
\inparg
{\it \noindent Fig.~1: The variation of the nature of the LSP (stau LSP 
below the line, selectron LSP above).}
\medskip 
\outparg

Once our results agree pretty with those of Ref.~\ADD, although 
our demarcation line is slightly lower, particularly for larger values of 
$\tan\beta$.

\newsec{The stop masses}
The bounds on the $\lambda''$ couplings are much weaker than for the 
$\lambda$ and $\lambda'$ couplings and in general are only set by perturbativity
of the top Yukawa coupling. The situation changes if a particular form is 
assumed for the quark mixing, such as mixing only in the up-quark or 
only in the down-quark sector. The bounds are particularly stringent in the
down-quark mixing case. Although, as described earlier, we assume
the no-mixing case, we expect our results to be qualitatively valid in the
general case and therefore we shall display our results up to the perturbativity
bound. We consider the dependence of the stop masses on $\lambda''_{323}$.
(In the no-mixing case it is clearly consistent to consider a single 
non-zero coupling at all scales.) The mass matrix for up-type quarks has no 
explicit dependence on the 
RPV couplings and so the dependence on $\lambda''_{323}$ is purely an implicit
effect due to the RG evolution. The stop masses are very sensitive to the 
value of the top mass; here as elsewhere in the paper we take 
$m_{\rm{top}}=174.3\GeV$.  
We see that the variation of the stop masses, 
especially the light one, on $\lambda''_{323}$ is considerable.
\bigskip
\epsfysize= 4in
\centerline{\epsfbox{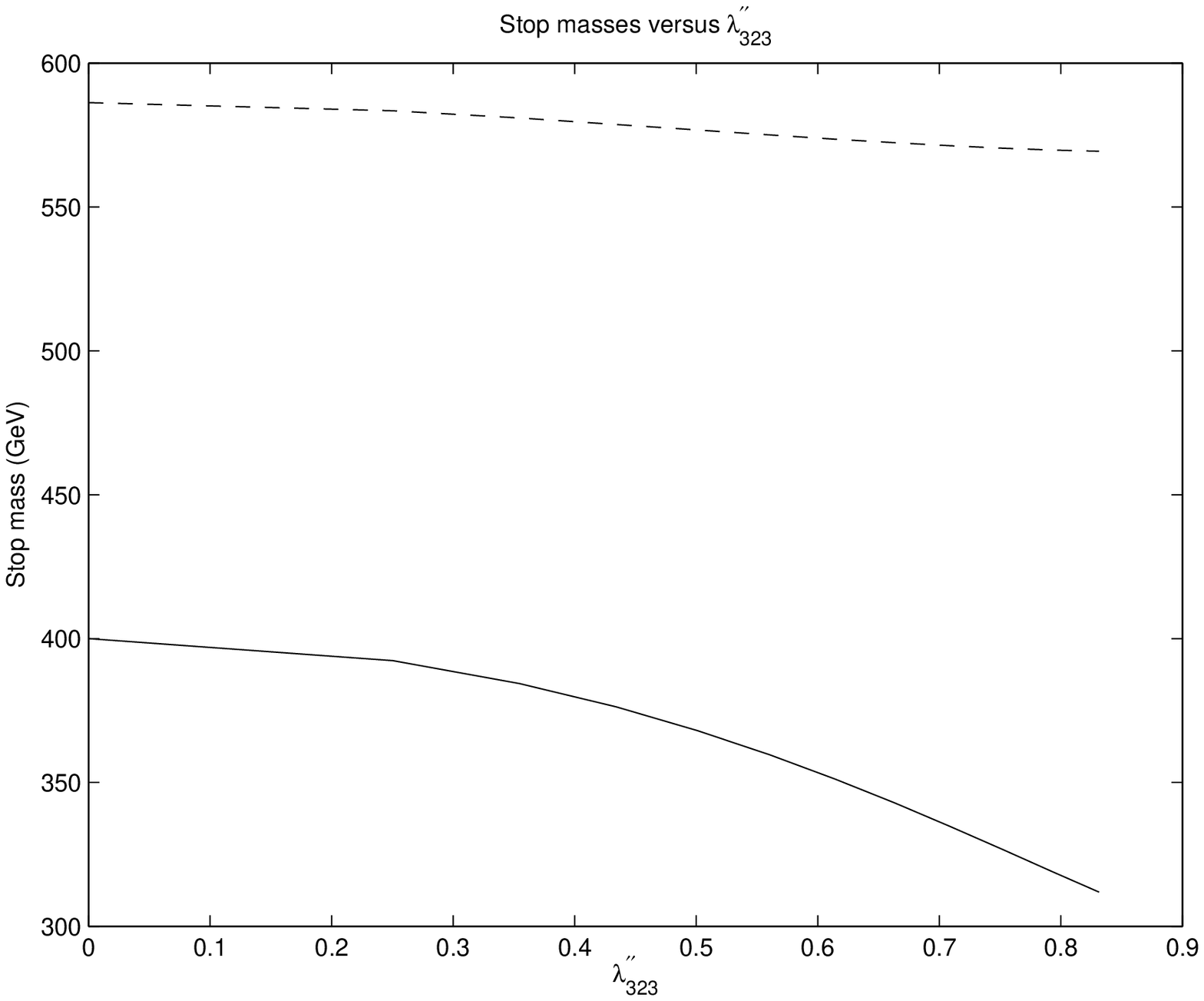}}
\inparg
{\it \noindent Fig.~2: Variation of the light stop mass with 
$\lambda''_{323}$.}
\medskip 
\outparg

\newsec{Conclusions}
We have analysed the effect of including the full set of two-loop
$\beta$-functions for R-parity violating couplings in a variety of scenarios.
Typically we find little difference between the effect of using the full
$\beta$-functions and that of using the one-loop $\beta$-functions plus
two-loop RPC corrections for RPC parameters; though as we see in Table 1 there
is quite a substantial difference between the bounds on RPV couplings obtained 
using the full two-loop $\beta$-functions and those obtained using 
the full one-loop $\beta$-functions--and of course it is desirable from the 
point of view of consistency to use the full set of $\beta$-functions.
In any event, we hope that future analysts will find the availability of the 
full set of $\beta$-functions
for the most general R-parity violating version of the MSSM to be a useful
resource\web. 

\bigskip\centerline{{\bf Acknowledgements}}\nobreak

DRTJ was supported by a PPARC Senior Fellowship, and a CERN Research
Associateship, and was visiting CERN while most of this work was done.
AK was supported by an Iranian Government Studentship. We are most grateful to 
Ben Allanach for providing us with detailed numerical output from the SOFTSUSY
program for purposes of comparison. 

\listrefs
\bye